\documentstyle[sprocl,epsf]{article}

\def\Journal#1#2#3#4{{#1} {\bf #2}, #3 (#4)}

\def\NPA{{\em Nucl. Phys.} A}

\def\PRL{\em Phys. Rev. Lett.}
\def\PRC{{\em Phys. Rev.} C}

\def\be{\begin{equation}}
\def\ee{\end{equation}}
\def\bea{\begin{eqnarray}}
\def\eea{\end{eqnarray}}
\makeatletter
\def\lsim{\mathrel{\mathpalette\subsim@align<}}
\def\gsim{\mathrel{\mathpalette\subsim@align>}}
\def\subsim@align#1#2{\lower.6ex\vbox{\baselineskip\z@skip\lineskip\z@
\ialign{$\m@th#1\hfil##\hfil$\crcr#2\crcr\sim\crcr}}}
\makeatother
\begin{document}

\title{Total and Parity-Projected Level Densities of Iron-Region
Nuclei by the Shell Model Monte Carlo Method}

\author{H. Nakada}

\address{Department of Physics, Chiba University,
Inage, Chiba 263, Japan\\E-mail: nakada@c.chiba-u.ac.jp}

\author{Y. Alhassid}

\address{Center for Theoretical Physics,
Yale University, New Haven, CT 06520, U.S.A.
\\E-mail: yoram@nst.physics.yale.edu}


\maketitle\abstracts{
Total and parity-projected level densities of  iron-region nuclei
are calculated microscopically by using  Monte Carlo methods for the
 nuclear shell model  in the complete $(pf+0g_{9/2})$-shell.
The calculated  total level density is found to be in good agreement
with the experimental level density. The Monte Carlo calculations offer a
significant improvement  over the thermal Hartree-Fock approximation.
Contrary to the Fermi gas model, it  is found that the level density
 has a significant parity-dependence in the neutron resonance region.
The systematics of  the level density parameters   (including shell effects)
in the iron region is presented.
}

\section{Introduction}
Neutron-capture reactions play
 an important role in nucleosynthesis,  e.g. in the $s$ and  $r$
 processes.
Their rates are strongly affected
by the corresponding nuclear level densities around the neutron resonance
region.
Most conventional calculations of the nuclear level density
are based on the Fermi gas model (e.g. the Bethe formula)~\cite{ref:BM1}.
A phenomenological modification is often adopted~\cite{ref:HWFZ}
where the excitation energy $E_x$ in the Bethe formula  is backshifted by
$\Delta$, giving a total level density of
\begin{eqnarray}\label{BBF}
 \rho (E_x) \approx
g  {{\sqrt\pi}\over{24}} a^{-\frac{1}{4}} (E_x - \Delta)^{-\frac{5}{4}}
 e^{2\sqrt{a (E_x - \Delta)}}
\end{eqnarray}
with $g=2$. The backshift $\Delta$ originates in pairing correlations
and shell effects, while
the parameter $a$ is determined by the single-particle level-density
at the Fermi energy. By adjusting the value of  $a$
for each nucleus, Eq.~(\ref{BBF}) describes well
a large volume of experimental data.
However, the  Fermi gas model grossly underestimates the value of $a$, and
consequently it is difficult to predict the level density to an accuracy
better than an order of magnitude. Much less is known about the
parity-dependence of the level density.

We have been studying the level densities of iron-region nuclei~\cite{ref:NA97}
by applying the recently proposed
shell model Monte Carlo (SMMC) method~\cite{ref:MCSM}
 within the the full $pf$- and $0g_{9/2}$-shell.
As a finite temperature method, it  is particularly suitable for
calculations of level densities (see below).
By introducing parity-projection methods
 for the auxiliary fields we are also able to calculate
parity-projected level densities.  We remark that the
 $(pf+0g_{9/2})$ model space is sufficiently  large
 to describe the important excitations around  the
neutron-resonance energies
in this mass region ($E_x\sim 5 - 15$~MeV).

\section{Calculations of level densities}\label{sec:comp}
We adopt an isoscalar Hamiltonian of the form~\cite{ref:NA97}
\begin{eqnarray}\label{2}
  H = \sum_a \epsilon_a \hat n_a + g_0 P^{(0,1)\dagger}\cdot \tilde P^{(0,1)}
     - \chi \sum_\lambda k_\lambda O^{(\lambda,0)}\cdot O^{(\lambda,0)} \;.
\end{eqnarray}
The single-particle energies $\epsilon_a$ are determined
from a Woods-Saxon potential $V$ plus spin-orbit interaction
with the parameters quoted in Ref.~\cite{ref:BM1}.
$P^{(0,1)\dagger}$ denotes the $T=1$ pair-creation operator
and $O^{(\lambda,T)}$ are multipole operators
with radial part of $dV/dr$~\cite{ref:ABDK}.
In the present calculations we include
multipole terms with  $\lambda=2,3$ and $4$.
 Core polarization effects are taken into account through
the use of renormalization factors $k_\lambda$. We use
 $k_2=2$, $k_3=1.5$ and $k_4=1$.
The pairing strength $g_0$ is determined from
 the experimental odd-even mass differences
for spherical  nuclei in the mass region $A=40-80$.
This Hamiltonian satisfies the modified sign rule~\cite{ref:ADK},
and therefore has a good Monte Carlo sign
for even-even nuclei. This enables us to perform accurate Monte Carlo
calculations.

In  SMMC the canonical  thermal  energy $E(\beta) \equiv\langle H\rangle_\beta$
 is calculated as a function of inverse temperature $\beta$.
Particle-number projection (for both protons and neutrons) is implemented
exactly~\cite{ref:MC-Nproj}.
 The canonical partition function $Z(\beta)$ is  determined
by a numerical integration of $E(\beta)$,
$\ln \left[ Z(\beta)/ Z(0)\right] =
 - \int_0^\beta d\beta^\prime E(\beta^\prime)$.
The level density $\rho(E)$ is then calculated
in the saddle-point approximation from
\begin{eqnarray}\label{7}
  \rho(E) &=& (2\pi \beta^{-2} C)^{-1/2}  e^{S} \;;
  \nonumber \\
 S(E) & = & \beta E + \ln Z(\beta) \;,
~~\beta^{-2}C(\beta)  = - dE / d\beta \;.
\end{eqnarray}
 $C$ is the heat capacity calculated
by numerical differentiation of $E(\beta)$.
To compare with experimental data, it is necessary to find
 $\rho$ as a function of the excitation energy. For that purpose we determine
the ground-state energy by extrapolating $E(\beta)$ to
$\beta\rightarrow\infty$.

To calculate the parity-dependence of the level density, we have introduced
 parity-projection techniques in SMMC~\cite{ref:NA97}.
Using the Hubbard-Stratonovich  representation
for $e^{-\beta H}$ and the projection operators
$P_\pm = (1 \pm P)/2$ ($P$ is the space reflection operator), we can  write
the projected energies  $E_\pm (\beta) \equiv { {\rm Tr} (H P_\pm e^{-\beta H} )
 / {\rm Tr} (P_\pm e^{-\beta H})}$ in the form
\begin{eqnarray} \label{E_pm}
 E_\pm (\beta)  = {\int D[\sigma]W(\sigma)  \left[ \langle H \rangle_\sigma \pm
\langle HP \rangle_\sigma \right]
\over
\int D[\sigma]W(\sigma) \left[  1 \pm \langle P \rangle_\sigma \right] }
\;.
\end{eqnarray}
The integration over the auxiliary fields $\sigma$
is performed with  the usual  Monte Carlo weight function
$W(\sigma) = G(\sigma) \zeta(\sigma)$,  where
$G$ is a gaussian factor  and $\zeta(\sigma) = {\rm Tr}~U_\sigma$ is the
 partition function  of  the non-interacting
propagator  $U_\sigma$~\cite{ref:MCSM}.
Both  $U_\sigma$ and $P$ can be represented
by  matrices in the single-particle space.
 $E_\pm (\beta)$ are then calculated through matrix algebra in this
single-particle space, similarly to the calculation of $E(\beta)$.
Once we find $E_\pm(\beta)$, the calculation of the projected
 densities $\rho_\pm(E)$ is analogous to that of the total level density.

\section{Results}
\subsection{$^{56}$Fe}
\begin{figure}
\epsfysize=5.0 cm
\centerline{\epsffile{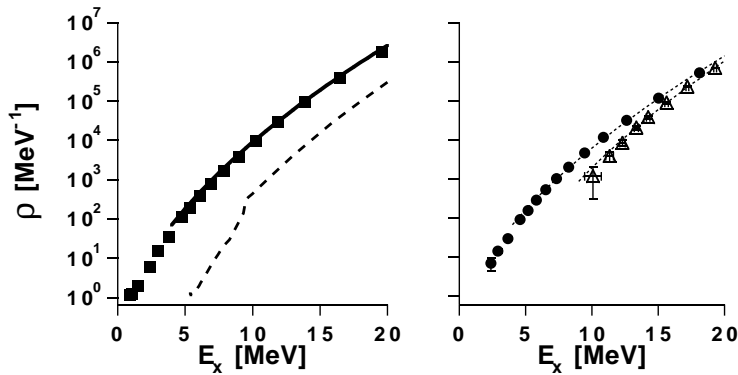}}
\vspace{2mm}
\caption{\protect\small Level densities of  $^{56}$Fe.
Left: Total level densities in  SMMC  (solid squares) and in the
HFA (dashed line). The solid line is the experimental
level density~\protect\cite{ref:LVH72}. Right: positive-parity
 (circles) and negative-parity
(triangles)  level densities in  SMMC.}
\label{fig:rho}
\end{figure}

Results for total
 level density in  $^{56}$Fe  are presented
in the left panel of  Fig.~\ref{fig:rho} as a function of $E_x$.
The SMMC results are the solid squares
 and include statistical errors
(often the errors are  too small to be visible).
Although it is difficult to measure the total level density directly,
it can be reconstructed from a few parameters
that are determined experimentally.
The solid line in Fig.~\ref{fig:rho} shows this experimental level density
 as determined from charged particle spectra~\cite{ref:LVH72}.
Our SMMC result is in good agreement
with the experimental results.
We have observed similar agreement (to within a factor of two)
 for other nuclei in the iron region.
Consequently, accurate level densities can be calculated
in the present approach.
In particular, we  extract the level density parameters $a$ and $\Delta$
via a fit  of our microscopically calculated level densities to Eq.~(\ref{BBF}).
Using the energy range $4.5~{\rm MeV}<E_x<20~{\rm MeV}$,
we obtain $a = 5.780\pm 0.055~{\rm MeV}^{-1}$
 and $\Delta = 1.560\pm 0.161~{\rm MeV}$ for $^{56}$Fe.
In Fig.~\ref{fig:rho} we also compare our SMMC results with those of
the thermal Hartree-Fock approximation (HFA)~\cite{ref:MFA} (dashed line).
The kink observed in the HFA level density
around $E_x \sim 9$~MeV is a signature of a shape phase transition,
but it is washed out
 in the SMMC due to strong two-body correlations.

The Fermi gas model  predicts
 equal positive- and negative-parity level densities at all energies.
However,  this is unrealistic in the neutron resonance regime of iron-region
nuclei where the excitation energy is comparable to
 the energy gap among major shells (i.e. the gap between
 the $pf$ and $g_{9/2}$ shells).
The SMMC  parity-projected level densities of $^{56}$Fe
are shown in the right panel of Fig.~\ref{fig:rho}.
They are well fitted to Eq.~(\ref{BBF}) with $g=1$,  but with
 parity-specific parameters $a_\pm$ and $\Delta_\pm$.
We find $a_+ = 5.611\pm 0.073 ~{\rm MeV}^{-1},~
 \Delta_+ = 0.550\pm 0.196 ~{\rm MeV}$ and
$a_- = 6.209\pm 0.625 ~{\rm MeV}^{-1},~
 \Delta_- = 3.172\pm 1.637 ~{\rm MeV}$.
Since negative-parity states in $^{56}$Fe are possible
only when $g_{9/2}$ is populated,
$\rho_-$ is lower than $\rho_+$ at low energies.
Thus the backshift $\Delta_-$ is larger than $\Delta_+$.
On the other hand, at high excitation energies
 positive- and negative-parity level densities are approximately equal,
resulting  in $a_- > a_+$.

\subsection{Systematics and shell effects}
\begin{figure}
\epsfysize=3.7 cm
\centerline{\epsffile{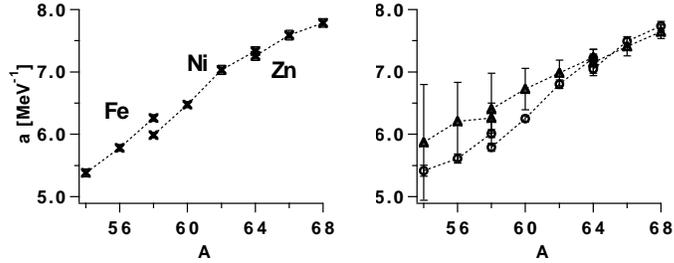}}
\vspace{2mm}
\caption{ \protect\small The single-particle level density parameter
$a$ as a function of $A$, for the total (left) and parity-projected
(right) level densities.  Points corresponding to isotopes with a 
given $Z$  are connected by dashed lines.
In the right panel, values of $a_+$ are shown by
 circles and values of $a_-$ by triangles. }
\label{fig:syst}
\end{figure}

We  discuss now the nucleus-dependence of the level density parameters
$a$ and $\Delta$ for  even-even nuclei in the $50<A<70$ region:
$^{54-58}$Fe, $^{58-64}$Ni and $^{64-68}$Zn.
Among them $^{54}$Fe and the  Ni isotopes have  $f_{7/2}$-subshell closure
for protons and/or neutrons ($Z$ or $N=28$).
Fig.~\ref{fig:syst} shows the calculated values of $a$.
It is interesting to see whether shell effects (at $Z$ or $N=28$)
can be observed in  the level density parameters.
We have found  enhancement of the backshift parameter  $\Delta$
at $Z$ or $N=28$~\cite{ref:NA98}.
However, no strong shell effects are observed in the
single-particle level density parameter $a$ and it
 increases smoothly as a function of $A$.
For  the parity-projected level densities we find that
$a_-$ increases as a function of $A$  more moderately than $a_+$.
This is because more excitations to the $g_{9/2}$ orbits  become  possible
(for fixed  $E_x$) as $A$ increases.
It would be interesting to investigate
how the parity-dependence affects the neutron-capture reaction rates.

\section{Conclusion}
We have used the auxiliary fields Monte Carlo method
to calculate the level density of  iron-region nuclei
in the complete $(pf+0g_{9/2})$-shell,
and found  good agreement with  experimental results.
The introduction of  a parity-projection technique in the SMMC
 allows us to study the parity-dependence of
the level density parameters.
The systematics of these parameters
 have been presented.

\section*{Acknowledgments}
This work was supported in part by  DOE grant
 DE-FG-0291-ER-40608,
and by the Ministry of Education, Science and Culture of Japan
(grant  08740190).

%This is where one places acknowledgments for funding
%bodies etc.  Note that there are no section numbers for
%the Acknowledgments, Appendix or References.


\section*{References}
\begin{thebibliography}{99}
\bibitem{ref:BM1} A. Bohr and B. R. Mottelson, {\it Nuclear Structure}
vol.~1 (Benjamin, New York, 1969).
\bibitem{ref:HWFZ} J. A. Holmes, {\it et al.},
 \Journal{\em Atom. Data and Nucl. Data Tables}{18}{305}{1976}.
\bibitem{ref:NA97} H. Nakada and Y. Alhassid,
\Journal{\PRL}{67}{2939}{1997}.
\bibitem{ref:MCSM} G. H. Lang, {\it et al.},
\Journal{\PRC}{48}{1518}{1993}.
\bibitem{ref:ABDK} Y. Alhassid, {\it et al.},
  \Journal{\PRL}{77}{1444}{1996}.
\bibitem{ref:ADK} Y. Alhassid, {\it et al.},
 \Journal{\PRL}{72}{613}{1994}.
\bibitem{ref:MC-Nproj}  W. E. Ormand, {\it et al.},
 \Journal{\PRC}{49}{1422}{1994}.
\bibitem{ref:NA98} H. Nakada and Y. Alhassid, in preparation.
\bibitem{ref:MFA} P. Quentin and H. Flocard,
\Journal{\em Ann. Rev. Nucl. Part. Sci.}{28}{523}{1978}.
\bibitem{ref:LVH72} C. C. Lu, L. C. Vaz and J. R. Huizenga,
\Journal{\NPA}{190}{229}{1972}.

\end{thebibliography}
\end{document}